\documentclass[prb,twocolumn,aps,showpacs,preprintnumbers,amsmath,amssymb,groupedaddress]{revtex4}

\usepackage{graphicx}
\usepackage{dcolumn}
\usepackage{bm}

\begin{document}

\preprint{PGF/ATB-1}

\title{Inward dispersion of the spin excitation spectrum of stripe-ordered
La$_{2}$NiO$_{4+\delta}$}

\author{P. G. Freeman}
\affiliation{Institut Laue-Langevin, BP 156, 38042 Grenoble Cedex
9, France}

\author{S. M. Hayden}
\affiliation{H. H. Wills Physics Laboratory, University of Bristol,
Bristol, BS8 1TL, United Kingdom}

\author{C. D. Frost}
\affiliation{ISIS Facility, Rutherford Appleton Laboratory,
Chilton, Didcot, OX11 0QX, United Kingdom}

\author{M. Enderle}
\affiliation{Institut Laue-Langevin, BP 156, 38042 Grenoble Cedex
9, France}

\author{D. X. Yao}
\affiliation{Department of Physics, Purdue University, West
Lafayette, Indiana 47907, USA}

\author{E. W. Carlson}
\affiliation{Department of Physics, Purdue University, West
Lafayette, Indiana 47907, USA}

\author{D. Prabhakaran}
\affiliation{Department of Physics, Oxford University, Oxford, OX1
3PU, United Kingdom }

\author{A. T. Boothroyd}
\email{a.boothroyd@physics.ox.ac.uk}\affiliation{Department of
Physics, Oxford University, Oxford, OX1 3PU, United Kingdom }

\date{\today}

\begin{abstract} Polarized- and unpolarized-neutron scattering measurements
of the spin excitation spectrum in the stripe-ordered phase of
La$_{2}$NiO$_{4+\delta}$ ($\delta \simeq 0.11$) are presented. At
low energies, the magnetic spectral weight is found to shift
anomalously towards the two-dimensional antiferromagnetic wave
vector, similar to the low-energy dispersions observed in cuprate
superconductors.  While spin-wave spectra in stripe phases can
exhibit an apparent inward dispersion, we find that the peak shifts
measured here cannot be accounted for by this effect. Possible
extensions of the model are discussed.

\end{abstract}

\maketitle

\section{Introduction}

Nano-periodic spin and charge correlations in the form of stripes
have featured prominently in the debate over the mechanism of
superconductivity in the layered cuprates \cite{Vojta-Archiv-2009}.
Of particular interest has been the interpretation of the spin
fluctuation spectrum measured by neutron scattering, which exhibits
a characteristic hour-glass shape in hole-doped cuprates
\cite{Arai-PRL-1999,Bourges-Science-2000,Hayden-Nature-2004,Tranquada-Nature-2004,Christensen-PRL-2004,Reznik-PRL-2004,Stock-PRB-2005,Vignolle-NatPhys-2007,Hinkov-NatPhys-2007,Matsuda-PRL-2008}.
The hour-glass dispersion is found in cuprates both with and without
static stripe order, and has recently been
observed\cite{Matsuda-PRL-2008} in a very lightly-doped, insulating,
cuprate in which the incommensurate spin modulation is diagonal,
i.e.\ at 45$^{\circ}$ to the Cu--O bonds, rather than parallel to
the bonds as found at higher doping. Theoretical approaches based on
stripes\cite{Tranquada-Nature-1995,Batista-PRB-2001,Vojta-PRL-2004,Uhrig-PRL-2004,Seibold-PRL-2005,Andersen-PRL-2005,Yao-PRL-2006}
generally succeed in reproducing the gross features of the spectrum,
but the low energy part remains a concern\cite{note-0}. These models
invariably predict cones of spin excitations emerging from four
equivalent incommensurate (IC) wavevectors (two in the case of
untwinned unidirectional stripes). Such spin-wave cones, however,
are not observed experimentally in the cuprates. Instead the spectra
show four IC peaks which disperse inwards with energy \emph{without
splitting} towards the ordering wavevector ${{\bf Q}_{\rm AF}}$ of
the parent antiferromagnet where they merge before dispersing apart
again at higher energies.

Insight into the role of stripes in the cuprates can be gained from
the structurally-related but insulating system
La$_{2-x}$Sr$_x$NiO$_{4+\delta}$, whose phase diagram exhibits
stable stripe order over a wide range of hole content
$n_h=x+2\delta$ without the complication of superconductivity
\cite{Hayden-PRL-1992,Chen-PRL-1993,Tranquada-PRL-1994,Yamada-PhysicaC-1994,Tranquada-PRB-1994,Yoshizawa-PRB-2000}.
Stripes in La$_{2-x}$Sr$_x$NiO$_{4+\delta}$ are aligned at
45$^{\circ}$ to the Ni--O bonds, like in lightly-doped
La$_{2-x}$Sr$_x$CuO$_{4}$, but as noted above, the characteristic
features of the magnetic dispersion in the cuprates is not dependent
on the alignment of the stripes. Detailed measurements of
La$_{2-x}$Sr$_x$NiO$_{4+\delta}$ have been reported for several
compositions close to $x=1/3$ with $\delta=0$
\cite{Bourges-PRL-2003,Boothroyd-PRB-2003,Boothroyd-PhysicaB-2004,Woo-PRB-2005}.
Like in the cuprates, the low energy spectra show four IC peaks, but
so far no evidence has been found for an inward dispersion. With the
exception of an unexplained peak near 25 meV
\cite{Boothroyd-PRB-2003,Boothroyd-PhysicaB-2004}, the spectra are
consistent with propagating spin-wave modes of ordered stripes.

In this paper we report neutron scattering measurements of the spin
excitation spectrum of La$_2$NiO$_{4.11}$. Previous studies of the
spin excitations in the nickelates have mostly been performed on
compounds whose stripe period is 1.5 to 2 times shorter than found
in stripe-ordered cuprates. In La$_2$NiO$_{4.11}$, however, the
stripe period is similar to that in the cuprates. Oxygen-doped
La$_{2}$NiO$_{4 + \delta}$ exhibits stripe order for $\delta \gtrsim
0.11$ ($n_h \gtrsim 0.22$)
\cite{Yamada-PhysicaC-1994,Tranquada-PRB-1994}. For $0 \leq \delta
\lesssim 0.11$,
commensurate antiferromagnetic (AFM) order of the parent phase
($\delta = 0$) is observed but with an ordering temperature that
decreases with $\delta$. The advantage of O-doping over Sr-doping is
that for a given hole concentration there is less disorder in the
stripes, possibly as a result of three-dimensional ordering of the
interstitial oxygen combined with a lack of cation disorder.

Our measurements reveal an anomalous inward dispersion of the
spectral weight in the energy range 10\,meV to $25$\,meV, while
above $\sim 25$\,meV the IC peaks are found to split into spin-wave
cones. This shows that the magnetic excitations of a stripe-ordered
antiferromagnet can display similar behavior at low energies to that
found in the hole-doped cuprates, although in La$_{2}$NiO$_{4 +
\delta}$ the complete hour-glass dispersion is not observed. Our
attempts to model the results quantitatively suggest that a
description that goes beyond linear spin-wave theory for ideal
stripes is required.

\section{Experimental details}

The single crystal of La$_{2}$NiO$_{4 + \delta}$ used here was grown
in Oxford by the floating-zone method \cite{Prabhakaran-JCG-2002}
and had a mass of 16\,g. The oxygen content determined by
thermogravimetric analysis of a specimen from the same boule was
$\delta = 0.11 \pm 0.01$.

Neutron scattering measurements were made on the MAPS time-of-flight
spectrometer at ISIS and on the IN8 and IN20 triple-axis
spectrometers at the ILL. A preliminary report of the MAPS data has
already been published \cite{Freeman-JMMM-2007}. On MAPS the crystal
was aligned with the $c$ axis parallel to the incident beam
direction. Spectra were recorded with several incident neutron
energies, and the scattering from a standard vanadium sample was
used to normalize the spectra and place them on an absolute
intensity scale. On IN8 and IN20 the crystal was mounted with the
$c$ axis and $[110]$ direction in the horizontal scattering plane
(we refer throughout to the tetragonal unit cell with lattice
parameters $a=3.85$\,{\AA} and $c=12.7$\,{\AA}), and the
spectrometers were configured for a fixed final neutron energy of
$E_{\rm f} = 14.7$\,meV with a graphite filter after the sample to
suppress higher orders.

Measurements on IN20 employed uniaxial polarization
analysis.\cite{Moon-PR-1969}. Polarized neutron diffraction (elastic
scattering) data were collected with three orthogonal orientations
of ${\bf P}$ relative to ${\bf Q}$: (1) ${\bf P}
\parallel {\bf Q}$, (2) ${\bf P} \perp {\bf Q}$ with ${\bf P}$ in the
scattering plane, and (3) ${\bf P} \perp {\bf Q}$ with ${\bf P}$
perpendicular to the scattering plane. The data were used to probe
the spin orientation in the magnetically-ordered phase. The
spin-flip (SF) and non-spin-flip (NSF) components of two magnetic
Bragg peaks were measured for each of the three orthogonal
directions of $\bf P$. Corrections were applied to compensate for
the imperfect neutron polarization. For the inelastic scattering
measurements the neutron spin polarization $\bf P$ was maintained
parallel to the scattering vector $\bf Q$ in order to separate
magnetic from non-magnetic scattering

\section{Results}
\begin{figure}[!ht] \begin{center}
\includegraphics
[width=6cm,bbllx=58,bblly=281,bburx=227,bbury=513,angle=0,clip=]
{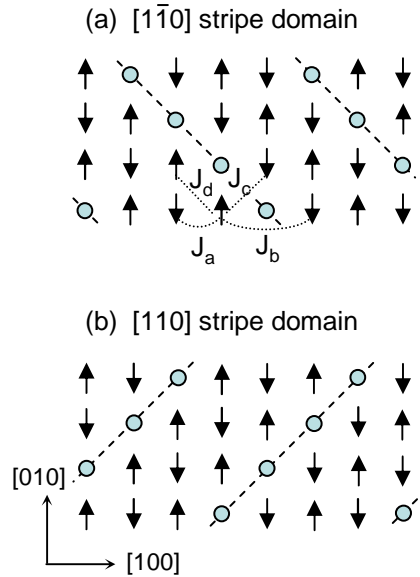} \caption{Model for ideal period--4 diagonal
stripes (DS4 model). Two equivalent domains are shown, with stripes
running parallel to the (a) $[1\bar{1}0]$ and (b) $[110]$ directions
of the square lattice. In relation to hole-doped nickelates, the
arrows show AFM ordered spins on Ni$^{2+}$ sites and the filled
circles represent doped holes assumed here to be localized on
Ni$^{3+}$ ions. Exchange interactions considered in the spin-wave
analysis are indicated. } \label{fig1}
\end{center} \end{figure}

Neutron diffraction measurements revealed patterns consistent with
diagonal stripe order (i.e.\ stripes oriented at 45$^{\circ}$ to the
Ni--O bonds) characterized by a fourfold group of magnetic
diffraction peaks at IC wavevectors ${\bf Q}_{\rm IC}={\bf Q}_{\rm
AF} \pm (\epsilon /2,\epsilon/2,0)$ and ${\bf Q}_{\rm AF} \pm
(-\epsilon /2,\epsilon/2,0)$, where $h$, $k$, $l$ are integers,
${\bf Q}_{\rm AF}=(h + 0.5,k+ 0.5,l)$ are antiferromagnetic (AF)
wavevectors, and $\epsilon=0.270\pm0.005$ is the incommensurability
\cite{note-1}. These peaks were observed below $T_{\rm N} \simeq
120$\,K, consistent with previous data
\cite{Yamada-PhysicaC-1994,Tranquada-PRB-1994}. The observation of
four satellites around ${\bf Q}_{\rm AF}$ is explained by the
presence of domains in which the charge stripes run along the two
equivalent diagonals of the square lattice, as shown schematically
in Fig.~\ref{fig1} for ideal period-4 stripes. The domain with
stripes parallel to the $[1\bar{1}0]$ direction gives rise to the
magnetic peaks displaced by $\pm(\epsilon /2,\epsilon/2,0)$ from
${\bf Q}_{\rm AF}$, while the $[110]$ stripe domain causes the peaks
displaced by $\pm(-\epsilon /2,\epsilon/2,0)$ from ${\bf Q}_{\rm
AF}$. The correlation lengths for the magnetic order were found to
be $\sim 50$\,{\AA} both parallel and perpendicular to the stripes
in the $ab$ plane, and $\sim 90$\,{\AA} in the $c$ direction. We
also observed the distinct sets of superlattice peaks associated
with interstitial oxygen ordering reported in Ref.
\onlinecite{Tranquada-PRB-1995}.

The polarized neutron diffraction measurements depend on the Fourier
component ${\bf M}({\bf Q})$ of the magnetic structure at the
ordering wavevector ${\bf Q}={\bf Q}_{\rm IC}$. Following the
approach described in Ref.~\onlinecite{Freeman-PRB-2002}, we used
the six measurements (SF and NSF components for three orientations
of $\bf P$ relative to $\bf Q$) at the magnetic Bragg peaks ${\bf
Q}_1 = (0.5-\epsilon /2,0.5-\epsilon /2,3)$ and ${\bf Q}_2 =
(0.5+\epsilon /2,0.5+\epsilon /2,1)$ to determine the intensities
associated with the components of ${\bf M}({{\bf Q}_{\rm IC}})$
along the $[110]$, $[1\bar{1}0]$ and $[001]$ directions. The data
collected at 2~K gave $I_{110}/I_{1\bar{1}0} = 0.062 \pm 0.004$ and
$I_{001}/(I_{110}+I_{1\bar{1}0}) = 0.004 \pm 0.004$. Within
experimental error these ratios did not vary with temperature
between 2~K and 60~K. For collinear magnetic order ${\bf M}({\bf
Q})$ is proportional to the ordered moment, and assuming this to be
the case here we find that the ordered moments lie in the $ab$ plane
to within an experimental uncertainty of 5~deg and make an in-plane
angle of $14.0 \pm 0.5$ deg to the stripe direction.

\begin{figure}
\begin{center}
\includegraphics
[width=8.0cm,bbllx=236,bblly=300,bburx=610, bbury=790,angle=0,clip=]
{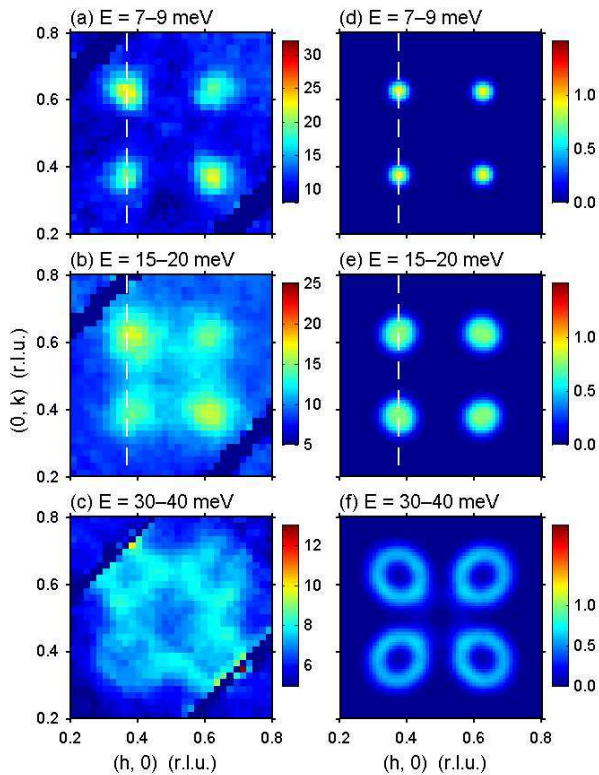} \caption{\label{fig2} (color online) Spin
excitation spectrum of La$_{2}$NiO$_{4.11}$ measured by inelastic
neutron scattering and calculated by linear spin-wave theory.
(a)--(c) Constant-energy slices averaged over the energy ranges
indicated above the figures. Vertical dashes mark the positions of
maximum intensity in the 7--9\,meV slices.  The data were collected
on the MAPS spectrometer with a sample temperature of 7\,K and with
incident neutron energies of 80\,meV. The intensity is in units of
mb\,sr$^{-1}$\,meV$^{-1}$\,f.u.$^{-1}$, where f.u. stands for
``formula unit" (of La$_{2}$NiO$_{4.11}$).
(e)--(g) Simulations based on the DS4 model shown in Fig.~\ref{fig1}
(Ref. \onlinecite{Carlson-PRB-2004}) with exchange constants $J_a =
28$\,meV, $J_b = 17$\,meV and $J_c = J_d = 0$\,meV, and 2D Gaussian
wavevector broadening with standard deviation 0.022\,r.l.u. The
intensity in the simulations is in arbitrary units.}
\end{center}
\end{figure}


\begin{figure}[!ht] \begin{center}
\includegraphics
[width=7cm,bbllx=56,bblly=200,bburx=530,bbury=604,angle=0,clip=]
{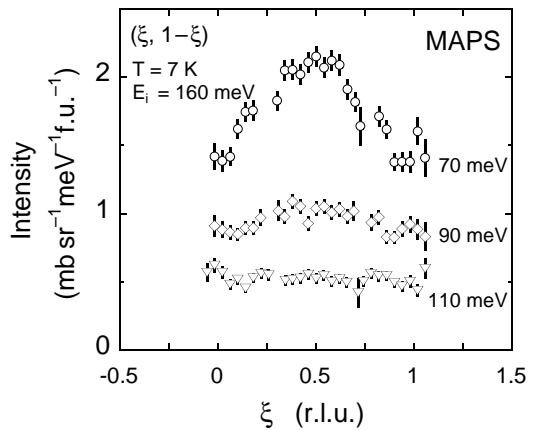} \caption{Constant-energy cuts through the
excitation spectrum of La$_{2}$NiO$_{4.11}$ parallel to the
$(\xi,-\xi)$ direction through $(0.5,0.5)$. No offset has been
applied to the scans at different energies.} \label{fig3}
\end{center} \end{figure}

We now turn to the magnetic excitation spectrum. Figs.\
\ref{fig2}(a)--(c) show constant-energy slices which illustrate the
most notable features of the observed spectrum. The data, which were
recorded on MAPS with unpolarized neutrons of incident energy
80\,meV and a sample temperature of 7\,K, have been averaged over
the indicated energy ranges and plotted as a function of the
in-plane scattering vector ${\bf Q}_{\rm 2D} = (h,k)$ in reciprocal
lattice units (r.l.u.) \cite{note-2}. The 7--9\,meV slice exhibits
four peaks at the magnetic ordering wavevectors. The intensities of
the peaks are unequal because the ordered moments make different
angles to $\bf Q$ in the two stripe domains.\cite{note-3} These four
peaks are also present in the 15--20 meV slice but have clearly
shifted inwards towards ${\bf Q}_ {\rm AF} = (0.5, 0.5)$.  Above
25\,meV these peaks evolve into rings as can be seen in the
30--40\,meV slice. The features just described are significantly
broader in wavevector than the experimental resolution, which is
approximately 0.05~r.l.u under the conditions used to collect the
data in Figs.\ \ref{fig2}(a)--(c). Above $\sim 50$\,meV the rings
strongly overlap and merge into a single broad peak centered on
${\bf Q}_ {\rm AF}$. As shown in Fig.~\ref{fig3}, the peak at ${\bf
Q}_ {\rm AF}$ decreases in intensity with increasing energy up to
$\sim 110$\,meV, above which no signal can be detected above the
background.

\begin{figure}[!ht] \begin{center}
\includegraphics
[width=8cm,bbllx=70,bblly=230,bburx=369,bbury=729,angle=0,clip=]
{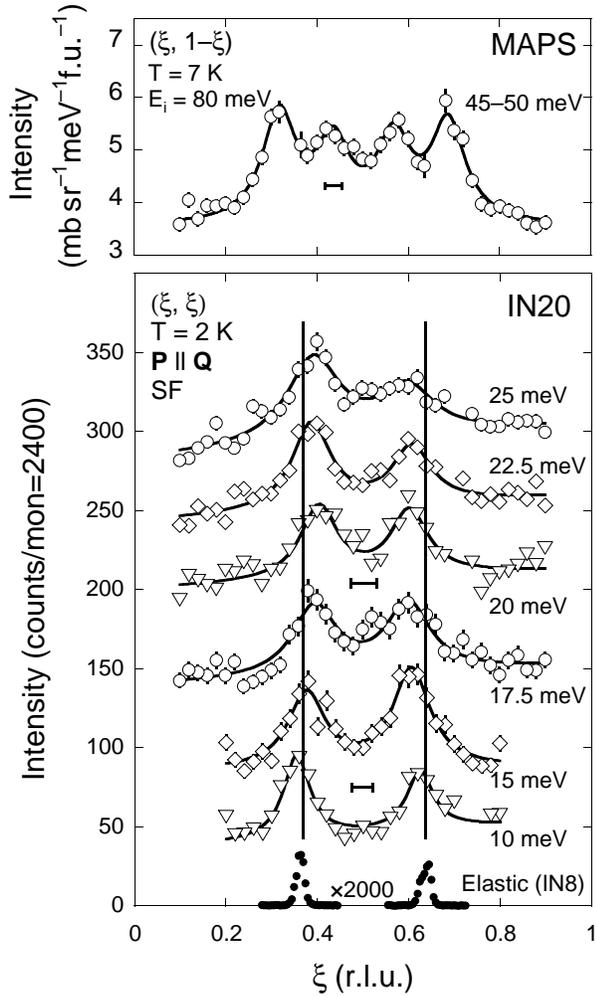} \caption[Energy dependence of magnetic
excitations] {Constant-energy scans through the spin excitation
spectrum of La$_{2}$NiO$_{4.11}$ in the direction perpendicular to
the stripes. The inelastic data in the lower panel were recorded on
IN20 employing neutron polarization analysis. The spin-flip (SF)
channel contains purely magnetic scattering. Vertical lines mark the
positions of the magnetic Bragg peaks, shown at the base of the
frame. The scans were measured in several different zones:
$(\xi,\xi,4)$ for 10 and 15\, meV, $(\xi+1,\xi+1,0)$ for 17.5, 20
and 25\,meV, and $(\xi+1,\xi+1,0.5)$ for 22.5\,meV. Each scan is
offset vertically by 50 counts. The solid lines are the results of
fits to two Lorentzian functions with equal widths. The upper panel
shows a cut through the MAPS data averaged over the energy range
45--50\,meV. The out-of-plane wavevector is $l \simeq 5$. The solid
line is a fit to four Lorentzians with equal peak widths. Horizontal
bars in both panels indicate the wavevector resolution. }
\label{fig4}
\end{center} \end{figure}

We performed additional measurements with neutron polarization
analysis to distinguish magnetic from non-magnetic scattering. Fig.\
\ref{fig4} (main panel) shows a series of constant-energy scans
along the $(\xi, \xi)$ direction measured on IN20 in the spin-flip
(SF) scattering channel. With {\bf P} $\parallel$ {\bf Q} the SF
scattering is purely magnetic. Up to 10\,meV the scans show peaks
centred at the same wavevectors as the magnetic Bragg peaks, which
are also shown in the figure, but above 10\,meV the peaks move
inwards and broaden compared with the resolution. Interestingly,
even at 10\,meV the peaks are about 50\% broader than the
experimental resolution.

These results confirm that the inward-dispersing peaks found here
are magnetic in origin. The upper panel of Fig.\ \ref{fig4} shows a
cut through the MAPS data along the direction perpendicular to the
stripes, averaged over the energy range 45--50\,meV. This
illustrates the splitting of the IC peaks at higher energies [see
also Fig.~\ref{fig2}(c)].

\begin{figure}[!ht] \begin{center}
\includegraphics
[width=7cm,bbllx=64,bblly=159,bburx=500,bbury=625,angle=0,clip=]
{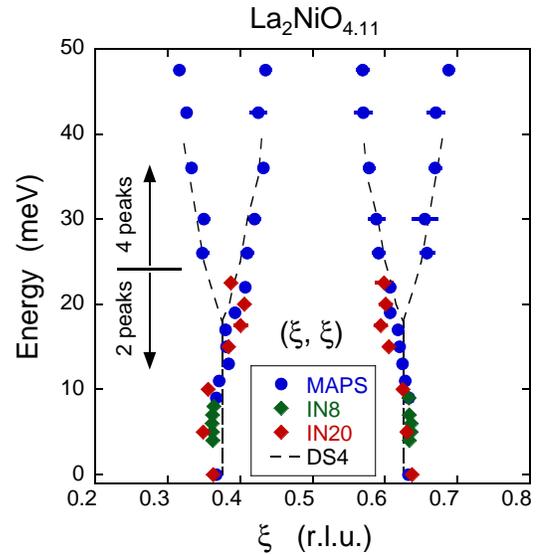} \caption[Fitted centres] {(color online)
Dispersion of magnetic excitations in La$_{2}$NiO$_{4.11}$. Data
points are peak centers from fits to a series of constant-energy
scans made on MAPS, IN20 and IN8. Data below (above) 25\,meV were
fitted to two (four) Lorentzian functions on a linear background.
The broken lines show the peak positions from the LSWT simulations
for the DS4 model [Fig.~\ref{fig1}].} \label{fig5}
\end{center} \end{figure}

To quantify the dispersion we fitted Lorentzian functions to the
peaks in a series of constant-energy scans including those shown in
Fig.~\ref{fig4}. The fitted peak centers are plotted in
Fig.~\ref{fig5}. Data from all three spectrometers are included in
the analysis and are consistent with one another. One can see
clearly how the peak intensity remains centered at the magnetic
ordering wavevectors up to 10\,meV, above which it disperses
inwards. At 20\,meV the incommensurability is 20--25\% less than the
static value. This analysis confirms that the shift observable in
the raw intensity data is intrinsic and not simply an artefact from
the overlap of the peaks. Above 25\,meV four peaks can be resolved
in the scans and the data were accordingly fitted to four Lorentzian
functions.
The resulting pair of ``V"-shaped dispersion curves are
asymmetric, with higher velocity on the inner branches.
This asymmetry partially counteracts the inward
shift in intensity at lower energies, but even in the highest energy
data (45--50\,meV) the pairs of spin-wave peaks are still shifted
towards ${\bf Q}_ {\rm AF}$ relative to the
magnetic Bragg peaks.

\begin{figure}[!ht] \begin{center}
\includegraphics
[width=7cm,bbllx=64,bblly=159,bburx=500,bbury=625,angle=0,clip=]
{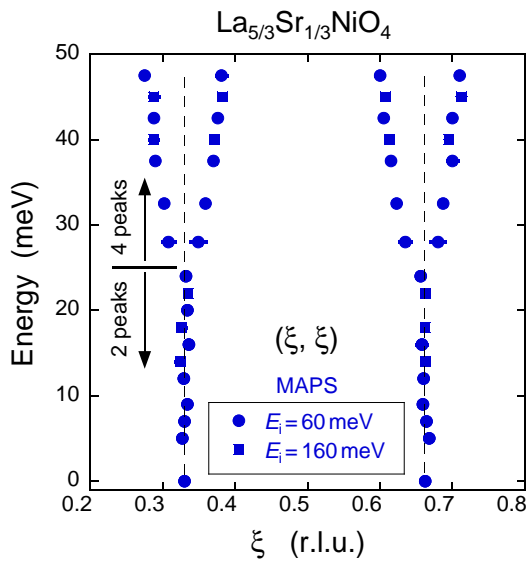} \caption{(color online) Dispersion of magnetic
excitations in La$_{5/3}$Sr$_{1/3}$NiO$_{4}$. The data are from the
study by Woo {\it et al.}\cite{Woo-PRB-2005} Points are peak centers
from fits to a series of constant-energy scans made on MAPS with
incident energies of 60\,meV and 160\,meV. Data below (above)
25\,meV were fitted to two (four) Lorentzian functions on a linear
background. The broken lines mark the magnetic Bragg peak
positions.} \label{fig6}
\end{center} \end{figure}

An obvious question to ask now is, do other stripe-ordered
nickelates exhibit similar features to those just described? The
question is answered for one other composition in Fig.~\ref{fig6}.
This figure shows the magnetic dispersion of
La$_{5/3}$Sr$_{1/3}$NiO$_{4}$, which has commensurate period-3
charge stripes ($\epsilon=1/3$). The dispersion was obtained from a
series of constant-energy cuts along the $(\xi, \xi)$ direction
through the MAPS data set of Woo {\it et al.}\cite{Woo-PRB-2005} As
before, a two- or four-Lorentzian line shape was used to obtain the
peak positions. To within experimental error the dispersion is
symmetric about the magnetic Bragg peak positions, in stark contrast
to the case of La$_{2}$NiO$_{4.11}$.

\section{Analysis and Discussion}

The data presented in the previous section show conclusively that
there is an inward shift towards ${\bf Q}_ {\rm AF}$ in the
intensity of the spin excitation spectrum of La$_{2}$NiO$_{4.11}$
for energies above $\sim$~10 meV, whereas there is no such shift for
the commensurate stripe-ordered compound
La$_{5/3}$Sr$_{1/3}$NiO$_{4}$. Since the spins in the stripe phase
are localized and ordered the dominant magnetic excitations are
expected to be spin precession waves, so a natural first step is to
see whether the results can be understood in the framework of linear
spin-wave theory.

Linear spin-wave theory (LSWT) has been used previously to
investigate the spin excitation spectrum of ideal commensurate
stripe structures \cite{Kruger-PRB-2003,Carlson-PRB-2004}. The
calculations assume a Heisenberg spin Hamiltonian, so charge
dynamics are included only insofar as they modify the strength of
the inter-stripe exchange couplings. To test whether LSWT can
describe the anomalous features of the spin excitation spectrum of
La$_{2}$NiO$_{4.11}$ we compared the data with LSWT predictions for
ideal site-centered period-4 diagonal charge stripes as shown in
Fig.~\ref{fig1}, dubbed DS4 in Ref.~\onlinecite{Carlson-PRB-2004},
for which the incommensurability $\epsilon=0.25$ is close to that
observed experimentally ($\epsilon\simeq0.27$). The minimal DS4
model has two exchange parameters, one ($J_a$) coupling
nearest-neighbor Ni$^{2+}$ spins within a stripe domain, and the
other ($J_b$) coupling Ni$^{2+}$ spins either side of a domain wall
in a straight line through a Ni$^{3+}$ site. The justification for
this model is that it provides a very good description of the
magnetic dispersion of La$_{5/3}$Sr$_{1/3}$NiO$_{4}$
\cite{Boothroyd-PRB-2003,Woo-PRB-2005}.

A detailed comparison with the model is not possible because the
experimental spectrum of La$_{2}$NiO$_{4.11}$ has considerable
intrinsic broadening and at high energies is complicated by the
twinning of the stripes. Instead, we estimate the parameters of the
model as follows. The magnetic dispersion of the $J_a$--$J_b$ DS4
model has a band width of $4J_a$ (we assume $S=1$). The experimental
magnon bandwidth is $110 \pm 10$\,meV, so $J_a = 28 \pm 3$\,meV. To
estimate $J_b$ we fitted the slope of the acoustic magnon branch
perpendicular to the stripes to the data above 25\,meV in Fig.\
\ref{fig5}. This gave $J_b = 17 \pm 3$\,meV. The values of $J_a$ and
$J_b$ agree closely with those determined experimentally for
La$_{5/3}$Sr$_{1/3}$NiO$_{4}$ ($J_a = 27.5 \pm 0.4$\,meV, $J_b =
13.6 \pm 0.3$\,meV \cite{Boothroyd-PRB-2003,Woo-PRB-2005}) and $J_a$
is in good agreement with that obtained for La$_{2}$NiO$_{4}$ ($J_a
= 31 \pm 0.7$\,meV \cite{Nakajima-JPSJ-1993}).

Figures \ref{fig2}(d)--(f) show simulated constant-energy slices
from the $J_a$--$J_b$ DS4 model alongside the corresponding data
from MAPS. The simulations reproduce most features of the data but,
crucially, they do not exhibit the inward dispersion of the
incommensurate peaks.

It has been shown that an apparent inward dispersion can be obtained
for striped spin structures within LSWT when $J_b \ll J_a$
\cite{Yao-PRL-2006}. In this case the inward dispersion is the
combined effect of broadening and a stronger intensity on the
surface of the spin-wave cone nearest ${\bf Q}_{\rm AF}$. A
reduction in $J_b/J_a$ is only possible within the constraints
imposed by the La$_{2}$NiO$_{4.11}$ data if we introduce additional
exchange parameters. However, the inclusion of diagonal
next-nearest-neighbor Ni--Ni couplings between ($J_c$) and within
($J_d$) stripe domains
did not enable us to reproduce the observed 20--25\% inwards
intensity shift within the DS4 model without departing from the
observed magnon band width of $\sim 110$\,meV and/or the data shown
in Fig.\ \ref{fig5}. We also performed simulations for period-4
bond-centred stripes but we could not reproduce the inwards
intensity shift with this model either.

Our results suggest that the magnetic dynamics of
La$_{2}$NiO$_{4.11}$ require a description beyond the simplest LSWT
of ideal stripes. Some of the following features may be needed:
First, the stripes in La$_{2}$NiO$_{4.11}$ are not commensurate but
have a period slightly less than four lattice spacings. A mixture of
$\sim 25$\% period-3 stripes and $\sim 75$\% period-4 stripes is
expected, and irregularities in the arrangement may explain the
observed broadening of the spin waves. Second, La$_{2}$NiO$_{4.11}$
is close to the border between stripe order and
La$_{2}$NiO$_{4}$-like AFM order
\cite{Yamada-PhysicaC-1994,Tranquada-PRB-1994}. Competition between
these order parameters may influence the magnetic spectrum. Third,
coupling between spin and charge degrees of freedom has so far been
neglected. Finally, we recall that an as-yet unexplained
resonance/gap-like feature appears in the the magnetic spectrum of
La$_{2-x}$Sr$_{x}$NiO$_{4}$ with $x \sim 1/3$
\cite{Boothroyd-PRB-2003,Boothroyd-PhysicaB-2004} in the same energy
range as the observed inward dispersion in La$_{2}$NiO$_{4.11}$. It
is possible that these anomalous features are related.

\section{Conclusion}

This work was motivated by the possibility that charge-stripe
correlations might be behind the hour-glass magnetic spectrum found
in the hole-doped cuprates. Our results show that one feature of the
cuprate spectrum, namely the inward dispersion at low energies, is
also found in an insulating nickelate with well-correlated but
incommensurate stripe order. This similarity is intriguing, but does
not necessarily imply that the inward magnetic dispersion has a
common origin in the two systems since there are important
differences between hole-doped cuprates and nickelates.
Nevertheless, our results do emphasize that the magnetic spectra of
stripe-ordered materials contain interesting features that need to
be understood better. Attempts to understand the inward dispersion
in the particular case of La$_{2}$NiO$_{4.11}$ will need to go
beyond the simplest model of ideal period-4 stripes used here.

\section*{ACKNOWLEDGMENTS}

A.T.B. is grateful to the Laboratory for Neutron Science at the Paul
Scherrer Institute for hospitality and support during an extended
visit in 2009. This work was supported by the Engineering \&
Physical Sciences Research Council of Great Britain, Research
Corporation, and US NSF Grant No. DMR 08-04748.


\begin{references}
\bibitem{note-1}
When $\epsilon=0.25$, stripes are commensurate with period 4.
\end{references}
\end{document}